\documentclass[pra,print,showpacs,superscriptaddress,twocolumn]{revtex4}
\usepackage[dvips]{graphics,color}
\usepackage{amsfonts}
\usepackage{amsmath}
\usepackage{amssymb}
\usepackage{graphicx}

\begin{document}

\title{Thermodynamics of spin-orbit-coupled Bose-Einstein condensates}
\author{Jinling Lian}
\thanks{myby1009@gmail.com}
\affiliation{Institute of Theoretical Physics, Shanxi University, Taiyuan 030006, P. R.
China}
\author{Yuanwei Zhang}
\thanks{zywznl@163.com}
\affiliation{Institute of Theoretical Physics, Shanxi University, Taiyuan 030006, P. R.
China}
\author{J. -Q. Liang}
\affiliation{Institute of Theoretical Physics, Shanxi University, Taiyuan 030006, P. R.
China}
\author{Jie Ma}
\affiliation{State Key Laboratory of Quantum Optics and Quantum Optics Devices, Laser
spectroscopy Laboratory, Shanxi University, Taiyuan 030006, P. R. China}
\author{Gang Chen}
\thanks{Corresponding author: chengang971@163.com}
\affiliation{State Key Laboratory of Quantum Optics and Quantum Optics Devices, Laser
spectroscopy Laboratory, Shanxi University, Taiyuan 030006, P. R. China}
\author{Suotang Jia}
\affiliation{State Key Laboratory of Quantum Optics and Quantum Optics Devices, Laser
spectroscopy Laboratory, Shanxi University, Taiyuan 030006, P. R. China}

\begin{abstract}
In this paper we develop a quantum field approach to reveal the
thermodynamic properties of the trapped BEC with the equal Rashba and
Dresselhaus spin-orbit couplings. In the experimentally-feasible regime, the
phase transition from the separate phase to the single minimum phase can be
well driven by the tunable temperature. Moreover, the critical temperature,
which is independent of the trapped potential, can be derived exactly. At
the critical point, the specific heat has a large jump and can be thus
regarded as a promising candidate to detect this temperature-driven phase
transition. In addition, we obtain the analytical expressions for the
specific heat and the entropy in the different phases. In the single minimum
phase, the specific heat as well as the entropy are governed only by the
Rabi frequency. However, in the separate phase with lower temperature, we
find that they are determined only by the strength of spin-orbit coupling.
Finally, the effect of the effective atom interaction is also addressed. In
the separate phase, this effective atom interaction affects dramatically on
the critical temperature and the corresponding thermodynamic properties.
\end{abstract}

\maketitle

\section{Introduction}

The spin orbit coupling (SOC), which describes the interaction between the
spin and orbit degrees of freedom of a particle, has not only generated many
interesting quantum phenomena in modern physics ranging from the nuclear
physics to condensed-matter physics, and but also become an important
resource for realizing fault-tolerant topological quantum computing \cite{CN}%
. By controlling the external lasers, the different kinds of SOCs have been
proposed to be simulated in the trapped Bose-Einstein condensates (BECs)
with the neutral atoms \cite{JR}. Especially, in recent experiment at NIST,
the equal Rashba and Dresselhaus SOCs has been realized successfully in the
ultracold $^{87}$Rb atoms by a couple of Raman lasers \cite{Lin}. Attributed
to this pioneer experiment, the investigation of SOC-driven BECs has
attracted much attentions. Moreover, rich many-body phenomena with no
analogy in condensed-matter physics (in BECs, all ultracold atoms can occupy
the same quantum state) have been predicted by considering the ground-state
properties \cite%
{Larson2009,WC,Wu,Yip2011,YP,YP2,SS,XXu,Zhu,Hu,ZhangDW,ZhangDW2,Ozawa,Deng,WeiZheng2012,Zhai2012,Shenoy2012,ZFXu,Fialko,JR2012}%
. For example, in the presence of the equal Rashba and Dresselhaus SOCs, the
BEC is made up of two non-orthogonal dressed atom spin states carrying
different momenta. Furthermore, the interaction between these spin states
are modified, driving a quantum phase transition from a spin-mixed state to
a phase-separated state \cite{TLH}. In fact, even if the effective atom
interaction governed by both the inter- and intra- spin interactions is not
taken into account, a quantum phase transition from a separate phase (SP) to
a single minimum phase (SMP) can also occur \cite{Zhang1,Yun}. In very
recent experiment, this new quantum phase transition has been observed by
measuring the amplitude ratio of spin and momentum oscillation \cite{ChenS}.

It has been known that quantum phase transitions governed by the
ground-state energies occur at absolute zero temperature \cite{Sachdev}.
However, it is unattainable experimentally due to the third law of
thermodynamics, i.e., any system must work at a finite temperature. Thus, it
is crucially important to investigate the thermodynamic properties to fully
understand the fundamental physics for a given system. For instance, in the
framework of finite-temperature theory, the system's real evaluation can be
described more accurately and some important physical quantities such as the
specific heat, the entropy and the free energy, which have no zero
temperature correspondence, can be explored. More importantly, some exotic
phenomena driven only by thermal fluctuations can be revealed \cite{Nagaosa}.

Motivated by the experimental developments and the third law of
thermodynamics, we, for the first time, develop a quantum field approach to
reveal the thermodynamic properties of the trapped BEC with the equal Rashba
and Dresselhaus SOCs. Our main results are given as follows: (I) In the
experimentally-feasible regime, the phase transition from the SP to the SMP
can be driven by the tunable temperature. Moreover, the corresponding
critical temperature is derived exactly and is independent of the trapped
potential. (II) We find that the specific heat has a large jump at the
critical temperature. This step behavior is quite different from that of the
atom population, which varies smoothly when crossing the critical point. It
implies that the temperature-driven phase transition can be well detected by
measuring the specific heat. (III) In the different phases, the analytical
expressions for the specific heat and the entropy are also given. In the
SMP, the specific heat as well as the entropy are governed only by the Rabi
frequency. However, in the SP, the strong SOC modifies the energy structure
and thus the thermodynamic statistics. At lower temperature, we find that
the specific heat and the entropy in such phase are determined only by the
SOC strength. (IV) Finally, the effect of the effective atom interaction is
also addressed. In the SMP, no collective excitations can be found in
SOC-driven BEC and thus the effective atom interaction does not affect the
thermodynamic properties. However, in the SP with SOC-induced macroscopic
excitations, this effective atom interaction affects dramatically on the
critical temperature as well as the other thermodynamic quantities. For
example, for the repulsive atom interaction, the critical temperature
decreases, and vice versa.

\begin{figure}[t]
\includegraphics[width = 0.7\linewidth]{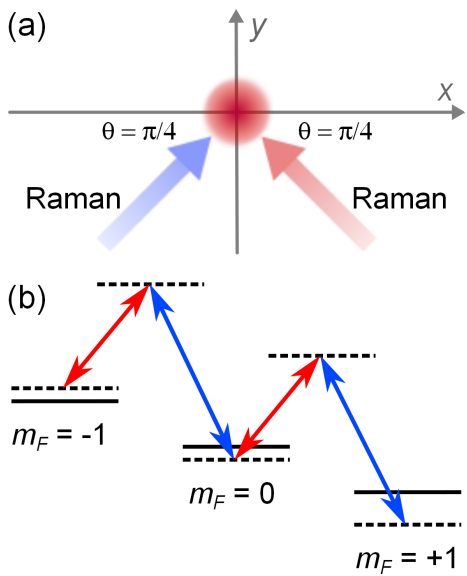}
\caption{(Color online) (a) The experiment setup for realizing the equal
Rashba and Dresselhaus SOCs in the trapped BEC at NIST \protect\cite{Lin}.
(b) The energy level structure of $^{87}$Rb atoms.}
\label{fig1}
\end{figure}

\section{Model and Hamiltonian}

Figure 1 shows the experimental scheme about how to create SOC in the
trapped BEC with the ultracold $^{87}$Rb atoms at NIST \cite{Lin}. In their
experiment, the BEC is trapped in the $xy$ plane through a strong
confinement with frequency $\omega _{z}$ along the $z$ direction. In the
large detuning $\Delta $, the momentum-sensitive coupling between two
hyperfine ground states $|F=1,m_{F}=-1\rangle (\left\vert \uparrow
\right\rangle )$ and $|F=1,m_{F}=0\rangle (\left\vert \downarrow
\right\rangle )$ is constructed by a pair of Raman lasers with Rabi
frequencies $\Omega _{1}$ and $\Omega _{2}$ incident at a $\pi /4$ angle
from the $x$ axis, as illustrated in Fig. 1(a). In the dressed-state basis $%
\left\vert \bar{\uparrow}\right\rangle =\exp \left( i\mathbf{k}_{1}\cdot
\mathbf{r}\right) \left\vert \uparrow \right\rangle $ and $\left\vert \bar{%
\downarrow}\right\rangle =\exp \left( i\mathbf{k}_{2}\cdot \mathbf{r}\right)
\left\vert \downarrow \right\rangle $, where $\mathbf{k}_{1}$ and $\mathbf{k}%
_{2}$ are the wavevectors of the Raman lasers, an effective SOC, which is
identical to the one-dimensional equal Rashba and Dresselhaus SOCs in
condensed-matter physics, can be achieved. Moreover, the corresponding
Hamiltonian with the atom-atom collision interaction can be written from the
coupled Gross-Pitaevskii equations as \cite{Zhang1}
\begin{equation}
H_{0}=\hbar \omega _{x}Na^{\dag }a+\hbar \Omega S_{x}-\gamma _{0}\sqrt{%
m\hbar \omega _{x}}i(a^{\dag }-a)S_{z}+\frac{\hbar q}{N}S_{z}^{2}.
\label{H1}
\end{equation}%
Here, $a^{\dag }a$ is a harmonic trap mode with $a=\sqrt{m\omega _{x}/2\hbar
}(x+ip_{x}/m\omega _{x})$ and $m$ being the atom mass. $S_{z}=(\Phi
_{\uparrow }^{\dagger }\Phi _{\uparrow }-\Phi _{\downarrow }^{\dagger }\Phi
_{\downarrow })/2$ reflects the experimentally-measurable population between
the different spin components. $\omega _{x}$ is the trapped frequency in the
$x$ direction. $\Omega =\Omega _{1}\Omega _{2}^{\ast }/\Delta $ is the
effective Rabi frequency. $\gamma _{0}=\sqrt{2}\hbar k_{L}/m$ with $\hbar
k_{L}=\sqrt{2}\pi \hbar /\lambda $ being the SOC strength, where $\lambda $
is the wavelength of the Raman laser. The effective atom interaction $q$ is
proportional to $N(g_{\uparrow \uparrow }+g_{\downarrow \downarrow
}-2g_{\uparrow \downarrow })$, where $g_{\uparrow \uparrow }=g_{\uparrow
\downarrow }=4\pi \hbar ^{2}N(c_{0}+c_{2})/(ma_{z})$ and $g_{\downarrow
\downarrow }=4\pi \hbar ^{2}Nc_{0}/ma_{z}$ are the inter- and intra- spin
interaction constants with $c_{0}$ and $c_{2}$ being the \textit{s}-wave
scattering lengths and $a_{z}=\sqrt{2\pi \hbar /m\omega _{z}}$. $N$ is the
total atom number.

If defining the number-dependent trapped frequency $\omega =N\omega _{x}$ and the effective SOC strength $\gamma =\sqrt{m}\gamma _{0}$,
Hamiltonian (\ref{H1}) can be rewritten in the rotating frame as ($\hbar =1$
henceforth)%
\begin{equation}
H=\omega a^{\dag }a+\Omega S_{z}+\frac{\gamma \sqrt{\omega }}{\sqrt{N}}%
(a^{\dag }+a)S_{x}+\frac{q}{N}S_{x}^{2}.  \label{H2}
\end{equation}%
In the following discussion, we focus mainly on Hamiltonian (\ref{H2}), in
which $\left\langle S_{x}\right\rangle \ $stands for the atom population.
Before proceeding, we estimate the relative parameters under current
experimental conditions \cite{Lin,ChenS,Zhang2}. In the experiment of NIST,
the tunable trapped frequency $\omega _{x}$ is of the order of 10 Hz, and
correspondingly, the number-dependent trapped frequency $\omega $ is of the order
of MHz for $N=1.8\times 10^{5}$. Parameter $\gamma ^{2}$ is of the order of
kHz for $\lambda =804.1$ nm. The effective Rabi frequency $\Omega $ can
range from zero to the order of MHz. In addition, since $c_{0}=100.86$ $%
a_{B} $ and $c_{2}=-0.46$ $a_{B}$ with $a_{B}$ being the Bohr radius, we
have $g_{\uparrow \uparrow }\simeq g_{\downarrow \downarrow }\simeq
g_{\uparrow \downarrow }$ and thus $q\simeq 0$. It means that the effective
atom interaction need not be taken into account in the NIST's experiment. It
should be pointed out that this effective atom interaction can be well
controlled through Feshbach resonance \cite{Chin}. Moreover, its magnitude
can reach the order of MHz near the Feshbach resonant point. Finally, we
will take $E_{\text{L}}=\hbar ^{2}k_{L}^{2}/2m$, which is of the order of
kHz, as the natural unit of the energy for simplicity.

\section{Thermodynamic equilibrium equation}

A key step to extract the thermodynamic properties of the SOC-driven BECs is
to obtain the partition function of Hamiltonian (\ref{H2}) \cite{Nagaosa}.
Here we develop a quantum field approach, i.e., an imaginary-time ($\tau =it$%
) functional path-integral technique, to arrive at the target. We first
rewrite the collective spin operators in the representation of the Grassmann
Fermi fields, namely, $S_{z}=\sum_{i=1}^{N}\left( \mu _{i}^{\dag }\mu
_{i}-\nu _{i}^{\dag }\nu _{i}\right) ,S_{+}=\sum_{i=1}^{N}\mu _{i}^{\dag
}\nu _{i}$, and $S_{-}=S_{+}^{\dag }$, where the Fermi operators $\mu
_{i}^{\dag }(\mu _{i})$ and $\nu _{i}^{\dag }(\nu _{i})$ satisfy the
anticommutator relations $\{\mu _{i}^{\dag },\mu _{j}\}=\{\nu _{i}^{\dag
},\nu _{j}\}=\delta _{ij}$. Furthermore, we transform the harmonic trap mode
$a^{\dag }(a)$ into a single mode bosonic field $\psi ^{\dag }(\psi )$. As a
consequence, the partition function is obtained by%
\begin{equation}
Z=\int \left[ d\eta (\tau )\right] \exp \left[ -A(\tau )\right] .
\label{PFC}
\end{equation}%
In Eq. (\ref{PFC}), $[d\eta (\tau )]=d[\psi ,\psi ^{\ast },\mu ,\mu ^{\ast
},\nu ,\nu ^{\ast }]$ is the path integral measure. The Euclidean action is
given by
\begin{equation}
A=\int_{0}^{\beta }d\tau \lbrack \psi ^{\ast }\partial _{\tau }\psi
+\sum_{i=1}^{N}(\mu _{i}^{\ast }\partial _{\tau }\mu _{i}+\nu _{i}^{\ast
}\partial _{\tau }\nu _{i})+H_{F}],  \label{AA}
\end{equation}%
where $\partial _{\tau }=\partial /\partial \tau $, $\beta =1/(k_{B}T)$ with
$k_{B}$ being the Boltzmann constant and $T$ being the system's temperature,
and
\begin{widetext}
\begin{equation}
H_{F}=\omega \psi ^{\ast }\psi +\Omega \sum_{i=1}^{N}\left( \mu _{i}^{\ast
}\mu _{i}-\nu _{i}^{\ast }\nu _{i}\right) +\sum_{i=1}^{N}[\frac{\gamma \sqrt{%
\omega }}{\sqrt{N}}\left( \psi +\psi ^{\ast }\right) \left( \mu _{i}^{\ast
}\nu _{i}+\nu _{i}^{\ast }\mu _{i}\right) +\frac{q}{N}\left( \mu _{i}^{\ast
}\mu _{i}-\nu _{i}^{\ast }\nu _{i}\right) ^{2}].  \label{HF}
\end{equation}
\end{widetext}Since Hamiltonian (\ref{H2}) has two degrees of freedom
including the spin and orbit cases, it is very difficult to directly discuss
the partition function to extract its fundamental thermodynamic properties.
The usual method is that we eliminate one degree of freedom by integrating
the Euclidean action $A$ \cite{Nagaosa}. Without the effective atom
interaction ($q=0$), the Euclidean action $A$ is a quadric term and the
corresponding integral is Gaussian. It means that in this case we can
integrate over the Grassmann Fermi fields and then obtain the partition
function of the bosonic mode. However, for nonzero $q$ ($q\neq 0$), the
integral in the Euclidean action $A$ is not Gaussian and thus the
corresponding integral is hard to be solved directly. Here we introduce an
auxiliary field $x$ to circumvent this difficult. Based on this auxiliary
field $x$, we have \cite{Aparicio2010}
\begin{widetext}
\begin{equation}
\exp [-\frac{q}{N}\sum_{i=1}^{N}\left( \mu _{i}^{\ast }\mu _{i}-\nu
_{i}^{\ast }\nu _{i}\right) ^{2}]\propto \int \left[ d\eta \right] \exp
\{\int_{0}^{\beta }d\tau \lbrack \frac{1}{q}x^{\ast }x-\sqrt{\frac{1}{N}}%
\sum_{i=1}^{N}(x+x^{\ast })(\mu _{i}^{\ast }\mu _{i}-\nu _{i}^{\ast }\nu
_{i})]\}.  \label{ME}
\end{equation}
\end{widetext}In analogy of the mean field approximation, the value of
auxiliary field $x$ determines $\left\langle S_{x}\right\rangle $, as will
be shown. Substituting the formula about the auxiliary field $x$ into the
Euclidean action $A$ yields%
\begin{equation}
A\left( \psi ,x\right) =A_{0}\left( \psi ,x\right) +\sum_{i}\int_{0}^{\beta
}d\tau \Phi _{i}^{\ast }G\left( \psi ,x\right) \Phi _{i},  \label{action}
\end{equation}%
where
\begin{equation}
\Phi _{i}=\left( \mu _{i}^{\ast },\nu _{i}^{\ast }\right) ^{T},  \label{WF}
\end{equation}%
\begin{equation}
A_{0}\left( \psi ,x\right) =\int_{0}^{\beta }d\tau \left[ \psi ^{\ast
}\left( \omega +\partial _{\tau }\right) \psi -x^{\ast }x\right] ,
\label{AZ}
\end{equation}%
and
\begin{equation}
G\left( \psi ,x\right) =\left[
\begin{array}{cc}
\partial \tau +\Omega & \mathcal{F}\left( \psi ,x\right) \\
\mathcal{F}\left( \psi ,x\right) & \partial \tau -\Omega%
\end{array}%
\right]  \label{GREEN}
\end{equation}%
with
\begin{equation}
\mathcal{F}\left( \psi ,x\right) =\gamma \sqrt{\frac{\omega }{N}}(\psi
^{\ast }+\psi )-\sqrt{\frac{1}{N}}(x^{\ast }+x).  \label{FF}
\end{equation}

For the effective Euclidean action $A$ in Eq. (\ref{action}), we can
integrate over the Grassmann Fermi fields, i.e., the degree of freedom for
the spin, and then obtain
\begin{equation}
A=N\int_{0}^{\beta }d\tau \left[ \Psi ^{\ast }\left( \omega +\partial _{\tau
}\right) \Psi -X^{\ast }X-\text{Tr}\ln G\right] ,  \label{FACT}
\end{equation}%
where $\Psi =\psi /\sqrt{N}$ and $X=x/\sqrt{N}$. Finally, by means of the
standard stationary phase approximation, namely, $\delta A/\delta \Psi
=\delta A/\delta \Psi ^{\ast }=0$ and $\delta A/\delta X=\delta A/\delta
X^{\ast }=0$, the required $\Psi $ and $X$, which play a crucial role in
determining thermodynamic properties of Hamiltonian (\ref{H2}), can be
obtained by
\begin{equation}
\left\{
\begin{array}{l}
\Psi =\Psi ^{\ast }=\frac{2}{\zeta }(\gamma ^{2}\Psi -\frac{\gamma }{\sqrt{%
\omega }}X)\tanh (\frac{\beta \zeta }{2}) \\
X=X^{\ast }=\frac{2}{\zeta }(\gamma q\sqrt{\omega }\Psi -qX)\tanh (\frac{%
\beta \zeta }{2})%
\end{array}%
,\right.  \label{EEQ}
\end{equation}%
where
\begin{equation}
\zeta =\sqrt{\Omega ^{2}+4(\gamma \sqrt{\omega }\Psi -X)^{2}}.  \label{PX}
\end{equation}%
It should be noticed that in the derivation of Eq. (\ref{EEQ}) we focus on
the constant path that $\Psi $\ is not influenced by $\tau $, namely, $%
\partial _{\tau }\Psi =0$ \cite{Nagaosa}. According to Eq. (\ref{EEQ}) we
have%
\begin{equation}
\Psi =\frac{2(\gamma ^{2}-q)}{\zeta }\tanh \left( \frac{\beta \zeta }{2}%
\right) \Psi .  \label{main1}
\end{equation}%
Equation (\ref{main1}) shows clearly that there exist a trivial solution $%
\Psi =\Psi ^{\ast }=0$, and the nontrivial solutions $\Psi =\Psi ^{\ast
}=\pm \Psi _{0}$ and $X_{0}=q\sqrt{\omega }\Psi _{0}/\gamma $ when $\gamma
\neq 0$. Moreover, these nontrivial solutions are governed by the nonlinear
equation $\zeta _{0}/[2\left( \gamma ^{2}-q\right) ]=\tanh (\beta \zeta
_{0}/2)$, where $\zeta _{0}=\sqrt{\Omega ^{2}+4(\gamma \sqrt{\omega }\Psi
_{0}-X_{0})^{2}}$. With the help of the stable condition at the equilibrium
points, we can obtain the required solutions of both $\Psi $ and $X$ and
thus reveal the thermodynamics of Hamiltonian (\ref{H2}) \cite{Nagaosa}.

\begin{figure}[t]
\includegraphics[width=7cm]{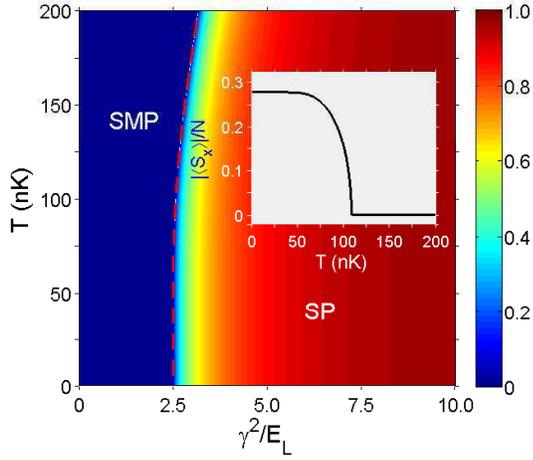}\newline
\caption{(Color online) The scaled atom population $|\langle S_{x}\rangle
|/N $ as the functions of the effective SOC strength $\protect\gamma $ and
the temperature $T$ without the effective atom interaction ($q=0$), where
the effective trapped frequency $\protect\omega \simeq 5.1\times 10^{3}E_{%
\text{L}}$ and the effective Rabi frequency $\Omega =5.0E_{\text{L}}$. The
red dashed line given by Eq. (\protect\ref{CRT}) determines the critical
boundary. Inset: The scaled atom population $|\langle S_{x}\rangle |/N$ as a
function of the temperature $T$ with the effective SOC strength $\protect%
\gamma ^{2}=2.6E_{\text{L}}$.}
\label{fig2}
\end{figure}

\section{Without effective atom interaction}

We first address the case of $q=0$, which has been realized at NIST \cite%
{Lin}. At zero temperature ($T=0$), $\tanh \left( \beta \zeta /2\right) =1$
and thus, Eq. (\ref{main1}) becomes $\Psi =2\gamma ^{2}\Psi /\zeta $, which
leads to solutions of $\Psi =\left\langle S_{x}\right\rangle =0$ for $\gamma
\leq \sqrt{\Omega /2}$ and $\Psi =\pm \sqrt{(4\gamma ^{4}-\Omega
^{2})/(4\omega \gamma ^{2})}$ and $\left\langle S_{x}\right\rangle =-\sqrt{%
1-\Omega ^{2}/(4\gamma ^{4})}$ for $\gamma \geq \sqrt{\Omega /2}$. These
zero-temperature solutions agree well with the direct numerical simulation
of the SOC-driven Gross-Pitaevskii equations \cite{Zhang1}. The nontrivial
variations of both $\Psi $ (atom momentum) and $\left\langle
S_{x}\right\rangle $ (atom population) show that a quantum phase transition
occurs by adjusting the effective SOC strength $\gamma $. Moreover, we can
call $\Psi =\left\langle S_{x}\right\rangle =0$ as the single minimum phase
(SMP) with no collective excitations, and $\Psi \neq 0$ and $\left\langle
S_{x}\right\rangle \neq 0$ as the separate phase (SP) with the macroscopic
excitations \cite{Yun}. With the increasing of the temperature $T$, the
order parameter $\Psi $ or $\left\langle S_{x}\right\rangle $ will be
destroyed by thermal fluctuation. In particular, when $\Psi
(T_{c})=\left\langle S_{x}\right\rangle (T_{c})=0$, the system enters into
the SMP from the SP. By means of $\Psi (T_{c})=0$, the critical temperature
can be obtained exactly by
\begin{equation}
T_{c}=\frac{\Omega }{2k_{B}\mathop{\rm arctanh}(\frac{\Omega }{2\gamma ^{2}})%
}.  \label{CRT}
\end{equation}%
Eq. (\ref{CRT}) shows that the phase transition from the SP to the SMP can
be driven by the tunable temperature. Moreover, the corresponding critical
temperature obtained exactly is independent of the trapped potential $\omega
$. When $\Omega =0.2E_{\text{L}}$ and $\gamma ^{2}=E_{\text{L}}$, the
critical temperature is evaluated as\textbf{\ }$T_{c}=84.9$ nK, which is
feasible in experiments about SOC-driven BECs.

\begin{figure}[t]
\includegraphics[width=8cm]{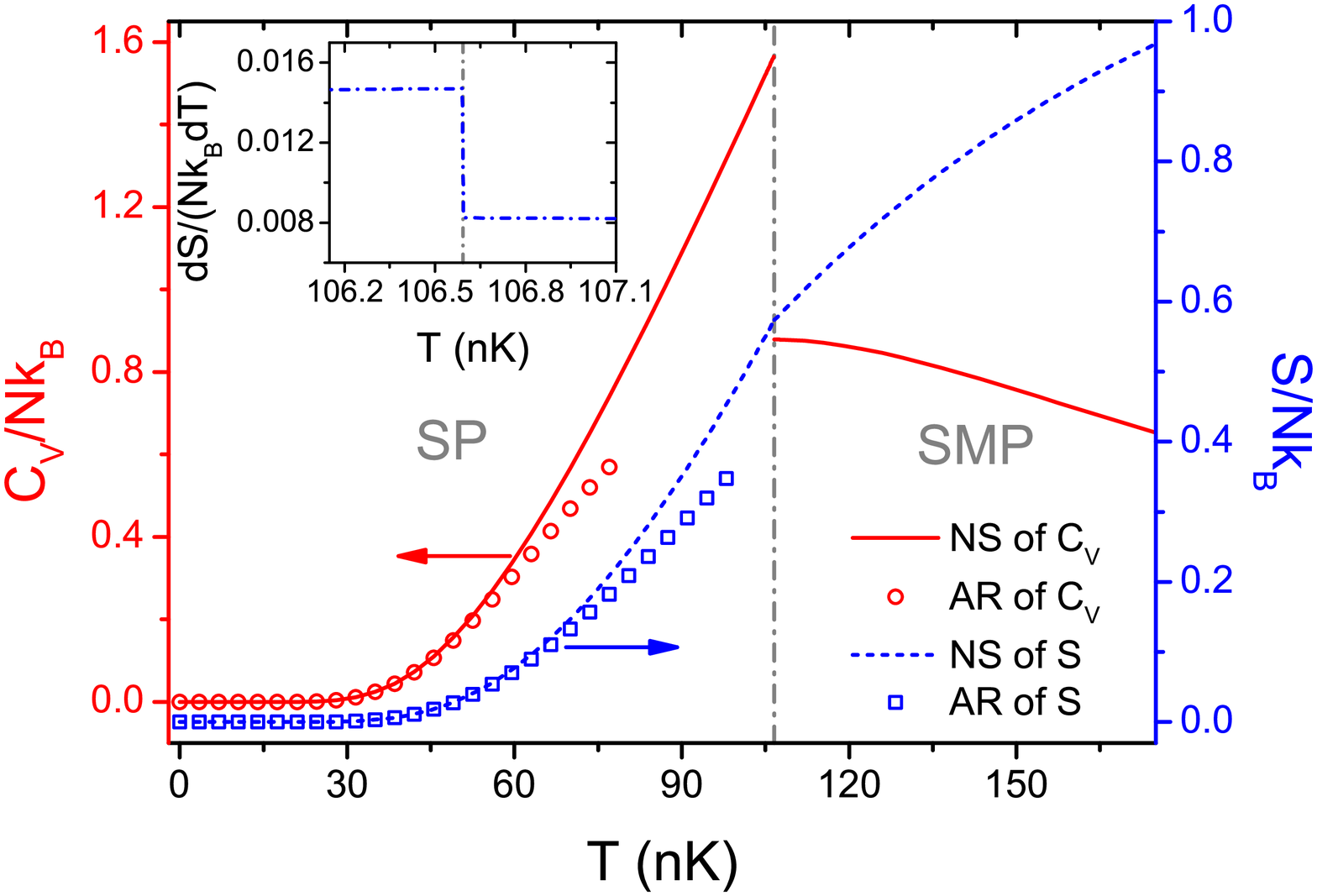}\newline
\caption{(Color online) The specific heat $C_{\text{v}}$ (Red lines) as well
as the entropy $S$ (Blue dashed lines) as a function of the temperature $T$
without the effective atom interaction ($q=0$). The plotted parameters are
given by $\protect\omega \simeq 5.1\times 10^{3}E_{\text{L}}$, $\Omega
=3.0E_{\text{L}}$, and $\protect\gamma ^{2}=1.8E_{\text{L}}$. These lines
and the circles (squares) denote the numerical simulation (NS) and the
analytical result (AR), respectively. Inset: The first-order derivative of $%
S/Nk_{B}$ versus the temperature $T$ (Blue dash-dotted line).}
\label{fig3}
\end{figure}

Having obtaining the critical temperature, we discuss the
experimentally-measurable atom population at finite temperature. In terms of
the thermodynamic equilibrium equation (\ref{main1}), the partition function
is given in the SMP with $\Psi (T)=0$ by
\begin{equation}
Z_{\text{SMP}}=\exp \{-N\beta \lbrack -\frac{2}{\beta }\ln (2\cosh (\frac{%
\beta \Omega }{2}))]\},  \label{PSMP}
\end{equation}%
whereas it becomes
\begin{equation}
Z_{\text{SP}}=2\exp \{-N\beta \lbrack \omega \Psi ^{2}-\frac{2}{\beta }\ln
(2\cosh (\frac{\beta \zeta }{2}))]\}  \label{PSP}
\end{equation}%
in the SP with $\Psi (T)\neq 0$. Thus, the atom population can be derived
from the formula%
\begin{equation}
\left\langle S_{x}\right\rangle (T)=\frac{\partial \left( \ln Z\right) }{%
-N\beta \partial \left( 2\gamma \sqrt{\omega }\Psi \right) }  \label{FOMA}
\end{equation}%
by
\begin{equation}
\left\langle S_{x}\right\rangle ^{\text{SMP}}(T)=0  \label{APT1}
\end{equation}%
in the SMP and
\begin{equation}
\left\langle S_{x}\right\rangle ^{\text{SP}}(T)=-\frac{\sqrt{\omega }}{%
\gamma }\Psi (T)  \label{APT2}
\end{equation}%
in the SP. In general, $\Psi (T)$ shoud be determined numerically by solving
the nonlinear equation (\ref{main1}). However, when $T=0$, Eqs. (\ref{APT1})
and (\ref{APT2}) reduce to the known analytical results \cite{Zhang1,Yun}.
In Fig. \ref{fig2}, we plot the scaled atom population $|\langle
S_{x}\rangle |/N$ as the functions of the effective SOC strength $\gamma $
and the temperature $T$. This figure shows that thermal fluctuations destroy
the collective excitations. As a result, the system finally enters into the
SMP from the SP.

\begin{figure}[t]
\includegraphics[width=7cm]{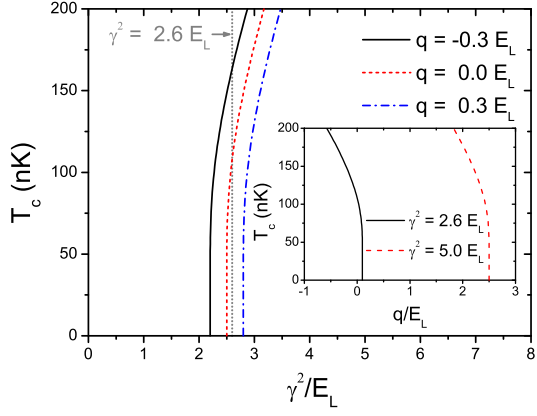}\newline
\caption{(Color online) The critical temperature $T_{c}$ as a function of
the effective SOC strength $\protect\gamma $ for the different effective
atom interactions $q=-0.3E_{\text{L}}$ (Black solid line), $q=0.0E_{\text{L}%
} $ (Red dashed line) and $q=0.3E_{\text{L}}$ (Blue dash-dotted line).
Inset: The critical temperature versus the effective atom interaction $q$
with the different effective SOC strength $\protect\gamma %
^{2}=2.6E_{\text{L}}$ (Black solid line) and $\protect\gamma ^{2}=5.0E_{%
\text{L}}$ (Red dotted line). In these figures, the other plotted parameters
are given by $\protect\omega \simeq 5.1\times 10^{3}E_{\text{L}}$ and $%
\Omega =5.0E_{\text{L}}$.}
\label{fig4}
\end{figure}

For a full understanding of the temperature-driven phase transition, it is
very important to discuss the thermodynamic quantities in the different
phases. Here we consider the specific heat per atom and the entropy per
atom. The other thermodynamic quantities can be calculated using the same
procedure. By means of the formula
\begin{equation}
C_{\text{V}}=\left( \frac{\partial U}{\partial T}\right) _{\text{V}}
\label{FOMC}
\end{equation}%
with $U=-N\frac{\partial }{\partial \beta }\ln Z$ being the total energy,
the specific heat per atom in the SMP is obtained exactly by
\begin{equation}
C_{\text{V}}^{\text{SMP}}=\frac{\Omega ^{2}}{2k_{B}T^{2}}{\mathop{\rm sech}}%
^{2}\left( \frac{\Omega }{2k_{B}T}\right) ,  \label{CVSB}
\end{equation}%
which is independent of both the trapped frequency $\omega $ and the
effective SOC strength $\gamma $. In the SP, the specific heat per atom is
evaluated as
\begin{widetext}
\begin{equation}
C_{\text{V}}^{\text{SP}}=\frac{1}{2k_{B}T^{2}}\left[\left( \zeta +\frac{\zeta
^{\prime }}{k_{B}T}\right) ^{2}{\mathop{\rm sech}}^{2}\left(\frac{\zeta }{2k_{B}T}\right)+2\left( \frac{\zeta ^{\prime \prime }}{k_{B}T}+2\zeta ^{\prime }\right)
\tanh \left(\frac{\zeta }{2k_{B}T}\right)-4\omega \left( \frac{\Psi ^{\prime 2}}{k_{B}T}
+2\Psi \Psi ^{\prime }+\frac{\Psi \Psi ^{\prime \prime }}{k_{B}T}\right) \right],
\label{CVSP}
\end{equation}
\end{widetext}where $\zeta =\sqrt{\Omega ^{2}+4\omega \gamma ^{2}\Psi ^{2}}$%
, $\zeta ^{\prime }=\partial \zeta /\partial \beta =4\gamma ^{2}\omega \Psi
\Psi ^{\prime }/\zeta $, and $\zeta ^{\prime \prime }=\partial ^{2}\zeta
/\partial \beta ^{2}=4\gamma ^{2}\omega \left( \Omega ^{2}\Psi ^{\prime
2}+\zeta ^{2}\Psi \Psi ^{\prime \prime }\right) /\zeta ^{3}$ with $\Psi
^{\prime }=\partial \Psi /\partial \beta =\zeta ^{2}/\left\{ 2\omega \Psi %
\left[ 1-2\beta \gamma ^{2}+\cosh \left( \beta \zeta \right) \right]
\right\} $ and $\Psi ^{\prime \prime }=\partial ^{2}\Psi /\partial \beta
^{2}=\{-16\gamma ^{4}\omega ^{2}[1-\beta \gamma ^{2}\mathop{\rm
sech}^{2}(\beta \zeta /2)]\Psi ^{2}\Psi ^{\prime 2}+\gamma ^{2}\zeta %
\mathop{\rm sech}^{2}(\beta \zeta /2)\tanh (\beta \zeta /2)(\zeta
^{2}+4\beta \gamma ^{2}\omega \Psi \Psi ^{\prime })^{2}+4\gamma ^{2}\omega
\zeta ^{2}[1-\beta \gamma ^{2}\mathop{\rm sech}^{2}(\beta \zeta /2)]\Psi
^{\prime 2}-8\gamma ^{4}\omega \zeta ^{2}\mathop{\rm sech}^{2}(\beta \zeta
/2)\Psi \Psi ^{\prime }\}/\{-4\gamma ^{2}\omega \zeta ^{2}[1-\beta \gamma
^{2}\mathop{\rm sech}^{2}(\beta \zeta /2)]\Psi \}$. The specific heat $C_{%
\text{V}}^{\text{SP}}$ implies that the strong SOC can modify the energy
structure of Hamiltonian (\ref{H2}) and thus the thermodynamic statistics,
as expected.

\begin{figure}[t]
\includegraphics[width=7cm]{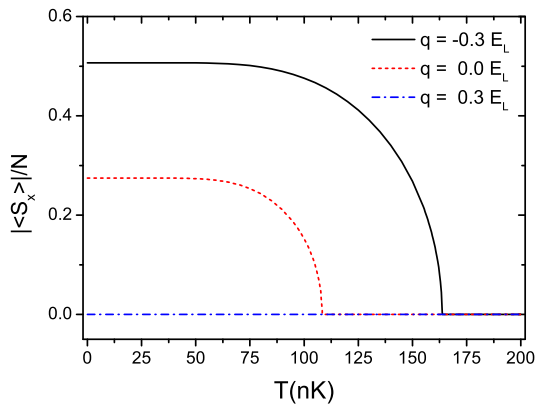}\newline
\caption{(Color online) The scaled atom population $\left\vert \left\langle
S_{x}\right\rangle \right\vert /N$ as a function of the temperature $T$ for
the different effective atom interactions $q=-0.3E_{\text{L}}$ (Black solid
line), $q=0.0E_{\text{L}}$ (Red dashed line) and $q=0.3E_{\text{L}}$ (Blue
dash-dotted line). The other plotted parameters are given by $\protect\omega%
\simeq5.1\times10^3 E_{\text{L}}$, $\Omega =5.0E_{\text{L}}$ and $\protect%
\gamma^{2}=2.6E_{\text{L}}$, respectively.}
\label{fig5}
\end{figure}

It is very hard to directly extract the fundamental properties of the
specific heat $C_{\text{V}}^{\text{SP}}$ from the complicate expression (\ref%
{CVSP}). In Fig. \ref{fig3}, we plot the specific heat $C_{\text{V}}$ (Red
lines) as a function of the temperature $T$. This figure shows two
interesting features. (I) The specific heat $C_{\text{V}}$ has a large jump
at the critical point $T_{c}$, separating the SP from the SMP. This step
behavior is quite different from that of the atom population $\left\langle
S_{x}\right\rangle $ in Fig. \ref{fig2}, which varies smoothly when crossing
the critical point. It implies that the temperature-driven phase transition
can be well detected by measuring the specific heat $C_{\text{V}}$. (II) At
lower temperature, we have approximately $\Psi ^{\prime }\simeq \Psi
^{\prime \prime }\simeq 0$. Thus, the specific heat in the SP can be
obtained analytically by
\begin{equation}
C_{\text{V}}^{\text{SP}}\simeq \frac{2\gamma ^{4}}{k_{B}T^{2}}{\mathop{\rm
sech}}^{2}\left( \frac{\gamma ^{2}}{k_{B}T}\right) ,  \label{CV}
\end{equation}%
which agrees well with the direct numerical simulations, as shown the Red
lines of Fig. \ref{fig3}. Eq. (\ref{CV}) shows, in contrast to the behavior
of the specific heat $C_{\text{V}}^{\text{SMP}}$, the specific heat $C_{%
\text{V}}^{\text{SP}}$ is governed only by the effective SOC strength $%
\gamma $ , i.e., it is independent of both the trapped potential $\omega $
and the effective Rabi frequency $\Omega $.

Another important thermodynamic quantity discussed in this paper is the
entropy, which can be derived from the formula
\begin{equation}
S=-\frac{\partial G}{\partial T}  \label{ENST}
\end{equation}
with $G=-k_{B}T\ln Z$ being the Gibbs function. In the SMP, the entropy per
atom is obtained exactly by
\begin{equation}
S^{\text{SMP}}=2k_{B}\ln \left[ 2\cosh \left( \frac{\Omega }{2k_{B}T}\right) %
\right] -\frac{\Omega }{T}\tanh \left( \frac{\Omega }{2k_{B}T}\right) ,
\label{ENT1}
\end{equation}%
whereas it becomes
\begin{widetext}
\begin{equation}
S^{\text{SP}}=\ln 2\frac{k_{B}}{{N}}+{2{k_{B}}\ln }\left[ {2\cosh }\left(
\frac{{\zeta }}{2k_{B}T}\right) \right] {-}\frac{{1}}{T}{\tanh }\left( {%
\frac{{\zeta }}{2k_{B}T}}\right) \left( {\zeta +}\frac{{{\zeta ^{\prime }}}}{%
k_{B}T}\right) {+}\frac{{2}}{k_{B}T^{2}}{\omega \Psi \Psi ^{\prime }}
\label{ENT2}
\end{equation}
\end{widetext}in the SP. At lower temperature, the entropy can be evaluated
approximately as
\begin{equation}
S^{\text{SP}}\!\!\simeq\! 2k_{B}\!\left\{ \!\frac{\ln 2}{2N}\!+\!\ln\!\! %
\left[ 2\cosh\!\! \left(\!\frac{\gamma ^{2}}{k_{B}T}\!\right)\! \right] \!-\!%
\frac{{\gamma ^{2}}}{k_{B}T}{\tanh \!\! \left(\!\frac{{\gamma^{2}}}{k_{B}T}%
\!\right)}\!\!\right\} .  \label{EntroyS}
\end{equation}
Similar to the behaviors of the specific heat $C_{\text{V}}$, the entropy $S$
in the SMP is governed only by the effective Rabi frequency $\Omega $,
whereas it is determined only by the effective SOC strength $\gamma$ in the
SP with lower temperature. However, its step behavior at the critical point $%
T_{c}$ is very small, as shown the Blue lines of Fig. \ref{fig3}.

\begin{figure}[t]
\includegraphics[width=7cm]{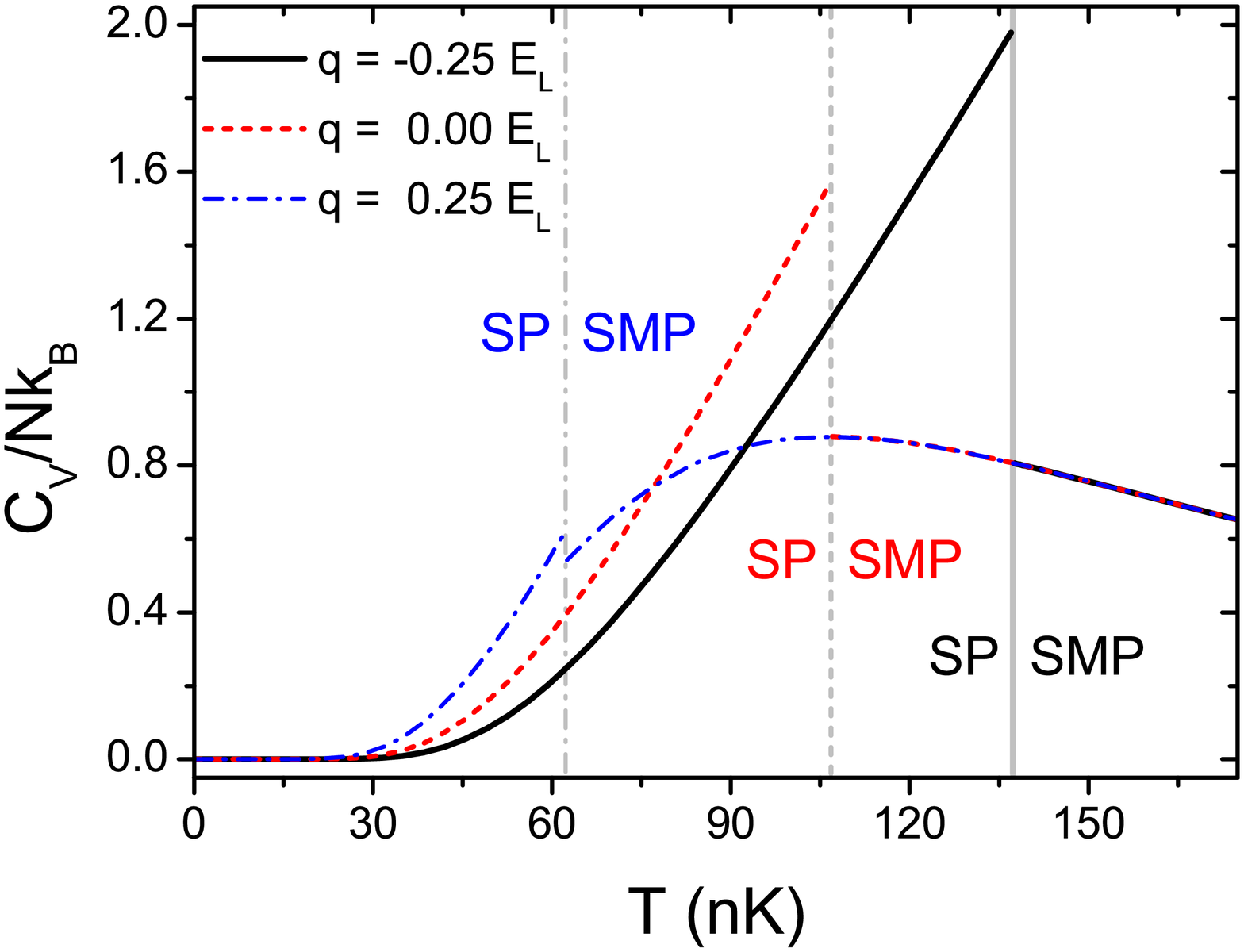}\newline
\caption{(Color online) The specific heat $C_{\text{V}}$ as a function of
the temperature $T$ for the different effective atom interactions $q=-0.25E_{%
\text{L}}$ (Black solid line), $q=0.00E_{\text{L}}$ (Red dashed line) and $%
q=0.25E_{\text{L}}$ (Blue dash-dotted line). The other plotted parameters
are given by $\protect\omega\simeq5.1\times10^3 E_{\text{L}}$, $\Omega
=3.0E_{\text{L}}$ and $\protect\gamma^{2}=1.8E_{\text{L}}$, respectively.
The three gray lines indicate the critical temperatures in the different
cases.}
\label{fig6}
\end{figure}

\section{With effective atom interaction}

In this section, we illustrate the effect induced by the effective atom
interaction, which is indeed controlled by the Feshbach resonant technique
in experiments. With the effective atom interaction ($q\neq 0$), the
critical temperature can be also obtained exactly by
\begin{equation}
T_{c}(q)=\frac{\Omega }{2k_{B}\mathop{\rm arctanh}\lbrack \frac{\Omega }{%
2(\gamma ^{2}-q)}]}.  \label{CTQ}
\end{equation}%
In Fig. \ref{fig4} we plot the critical temperature $T_{c}$ as a function of
the effective SOC strength $\gamma $ for the different effective atom
interactions. This figure shows clearly that for the attractive interaction (%
$q<0$), the critical temperature $T_{c}$ increases for a fixed SOC strength $%
\gamma $, and vice versa. It means that in experiments we can manipulate the
effective atom interaction $q$ to arrive at the experimentally-required
critical temperature $T_{c}$, where the thermodynamic phase transition from
the SP to the SMP occurs.

On the other hand, in the SMP, no collective excitations can be found in
SOC-driven BEC and thus the effective atom interaction does not change the
energy structure. It implies that the partition function in the SMP is the
same as Eq. (\ref{PSMP}), i.e., it is independent of the effective SOC
strength $\gamma $, the trapped frequency $\omega $ and the effective atom
interaction $q$. However, in the SP the strong SOC leads to the system's
collective excitations with nonzero atom population. As a result, the term $%
\frac{q}{N}S_{x}^{2}$ in Hamiltonian (\ref{H2}) plays a crucial role in
system's energy structure and thus the thermodynamic statistics. In such
case, the partition function becomes
\begin{equation}
Z_{\text{SP}}(q)=2\exp \{-N\beta \lbrack \frac{\omega \delta \Psi ^{2}}{%
\gamma ^{2}}-\frac{2}{\beta }\ln [2\cosh (\frac{\beta \zeta }{2})]]\},
\label{PSPQ}
\end{equation}%
where $\delta (q)=\gamma ^{2}-q$, $\Psi $ and $\zeta $ can be obtained from
the nonlinear equation (\ref{main1}) with the effective atom interaction $q$.

Based on the obtained partition function, the experimentally-measurable atom
population is given by $\left\langle S_{x}\right\rangle ^{\text{SMP}}(q,T)=0$
and $\left\langle S_{x}\right\rangle ^{\text{SP}}(q,T)=-\frac{\sqrt{\omega }%
}{\gamma }\Psi (q,T)$, which are plotted in Fig. \ref{fig5}. It is clearly
that in the presence of the attractive interaction ($q<0$), the critical
temperature increases and the system is more inclined to locate at the SP,
and vice versa. This conclusion is identical to the result of Fig. \ref{fig4}
(along the gray dotted line $\gamma ^{2}=2.6E_{\text{L}}$ there). In
addition, in the SMP, the specific heat $C_{\text{V}}$ and the entropy $S$
are also identical to the results of $q=0$. Whereas, in the SP they become $%
C_{\text{V}}^{\text{SP}}(q)=\frac{k_{B}\beta ^{2}}{2}[\left( \zeta +\beta
\zeta ^{\prime }\right) ^{2}{\mathop{\rm sech}}^{2}\left( \beta \zeta
/2\right) +2\left( 2\zeta ^{\prime }+\beta \zeta ^{\prime \prime }\right)
\tanh \left( \beta \zeta /2\right) -4\omega \left( 1-q/\gamma ^{2}\right)
(2\Psi \Psi ^{\prime }+\beta \Psi ^{\prime 2}+\beta \Psi \Psi ^{\prime
\prime })]$ and $S^{\text{SP}}(q)=k_{B}\{\ln 2/N+2\ln [2\cosh (\beta \zeta
/2)]-\beta (\zeta +\beta \zeta ^{\prime })\tanh (\beta \zeta /2)+2\beta
^{2}\omega (1-q/\gamma ^{2})\Psi \Psi ^{\prime }\}$, where $\zeta (q)=\sqrt{%
\Omega ^{2}+4\omega \eta ^{2}\Psi ^{2}}$, $\zeta ^{\prime }$ $(q)=4\eta
^{2}\omega \Psi \Psi ^{\prime }/\zeta $, $\Psi ^{\prime }(q)=\gamma
^{2}\zeta ^{2}/\{2\delta \omega \left[ 1-2\beta \delta +\cosh \left( \beta
\zeta \right) \right] \Psi \}$, and $\Psi ^{\prime \prime }(q)=\{-16\eta
^{4}\omega ^{2}[1-\beta \delta {\mathop{\rm
sech}}^{2}\left( \beta \zeta /2\right) ]\Psi ^{2}\Psi ^{\prime 2}+\delta
\zeta {\mathop{\rm sech}}^{2}\left( \beta \zeta /2\right) \tanh \left( \beta
\zeta /2\right) [\zeta ^{2}+4\beta \eta ^{2}\omega \Psi \Psi ^{^{\prime
}}]^{2}+4\eta ^{2}\omega \zeta ^{2}[1-\beta \delta {\mathop{\rm sech}}%
^{2}\left( \beta \zeta /2\right) ]\Psi ^{\prime 2}-8\gamma \eta ^{3}\omega
\zeta ^{2}{\mathop{\rm sech}}^{2}\left( \beta \zeta /2\right) \Psi \Psi
^{\prime }\}/\{-4\omega \eta ^{2}\zeta ^{2}[1-\beta \delta {\mathop{\rm sech}%
}^{2}\left( \beta \zeta /2\right) ]\Psi \}$ with $\eta (q)=\delta (q)/\gamma
=\gamma -q/\gamma $. When $T\ll T_{c}$, they reduce to the forms
\begin{equation}
C_{\text{V}}^{\text{SP}}(q)\simeq \frac{2\delta ^{2}}{k_{B}T^{2}}{%
\mathop{\rm sech}}^{2}\left( \Lambda \right) ,  \label{CVQ}
\end{equation}%
and
\begin{equation}
S^{\text{SP}}(q)\simeq 2k_{B}\left\{ \frac{\ln 2}{2N}+\ln \left[ 2\cosh
\left( \Lambda \right) \right] -\Lambda \tanh \left( \Lambda \right)
\right\} ,  \label{ENTQ}
\end{equation}%
where $\Lambda (q)=\delta (q) /k_{B}T$. In Fig. \ref{fig6}, the specific
heat $C_{\text{V}}$ as a function of the temperature $T$ for the different
effective atom interactions is plotted. This figure shows that the
fundamental properties of temperature-driven phase transition remain in the
framework of the effective atom interaction. However, for the repulsive
interaction ($q>0$), the critical temperature $T_{c}$ can decrease.
Moreover, the step amplitude at the critical point also decreases. For the
attractive interaction ($q<0$), the opposite results exist. The similar
behaviors of the entropy $S$ can be also found.

\section{Conclusions and Remarks}

In summary, we have explored the thermodynamic properties of the trapped BEC
with the equal Rashba and Dresselhaus SOCs, which has been realized in
experiments. The thermodynamic phase transition from the SP to the SMP as
well as the critical temperature has been revealed. We have also discussed
the important thermodynamic quantities such as the specific heat and the
entropy and obtained their analytical expressions in the different phases.
At the critical point, the specific heat has a large jump and can be thus
regarded as a promising physical quantity to detect this temperature-driven
phase transition. Finally, we have illustrated the effect of the effective
atom interaction, which can be well controlled by the
experimentally-feasible Feshbach resonant technique. Especially, we have
found that in the SP this effective atom interaction affects dramatically on
the critical temperature and the corresponding thermodynamic properties for
the SOC-driven BEC. Before ending up this paper, we briefly make two
remarks. Firstly, our analysis is mainly based on the imaginary-time
functional path-integral approach, in which the fluctuation of the
space-dependent physical quantity (such as the density) is usually
\textquotedblleft hidden\textquotedblright\ or \textquotedblleft averaged
out\textquotedblright\ \cite{Nagaosa}. As a consequence, the important
stripe phase predicted before \cite{WC,TLH,Yun} cannot be distinguished from
the SP effectively. Secondly, without SOC ($\gamma =0$), Hamiltonian (\ref%
{H2}) turns into $H=\Omega S_{z}+\frac{q}{N}S_{x}^{2}$. In this case, the
relation $X_{0}=q\sqrt{\omega }\Psi _{0}/\gamma $ becomes invalid. However,
we can use the same procedure (introducing the auxiliary field in
path-integral technique) to discuss the corresponding thermodynamics.

\section{Acknowledgements}

We thank Prof. Chuanwei Zhang and Dr. Yongping Zhang for their helpful
discussions. This work was supported partly by the 973 program under Grant
No. 2012CB921603; the NNSFC under Grants No. 10934004, No. 60978018, No.
11074154, No. 11075099, No. 61008012, No. 11275118, and No. 61275211; NNSFC
Project for Excellent Research Team under Grant No. 61121064; and
International Science and Technology Cooperation Program of China under
Grant No.2001DFA12490.

\end{document}